# Proficient Pair of Replacement Algorithms on L1 and L2 Cache for Merge Sort

Richa Gupta, Sanjiv Tokekar

**Abstract**— Memory hierarchy is used to compete the processors speed. Cache memory is the fast memory which is used to conduit the speed difference of memory and processor. The access patterns of Level 1 cache (L1) and Level 2 cache (L2) are different, when CPU not gets the desired data in L1 then it accesses L2. Thus the replacement algorithm which works efficiently on L1 may not be as efficient on L2. Similarly various applications such as Matrix Multiplication, Web, Fast Fourier Transform (FFT) etc will have varying access pattern. Thus same replacement algorithm for all types of application may not be efficient. This paper works for getting an efficient pair of replacement algorithm on L1 and L2 for the algorithm Merge Sort. With the memory reference string of Merge Sort, we have analyzed the behavior of various existing replacement algorithms on L1. The existing replacement algorithms which are taken into consideration are: Least Recently Used (LRU), Least Frequently Used (LFU) and First In First Out (FIFO). After Analyzing the memory reference pattern of Merge Sort, we have proposed a Partition Based Replacement algorithm (PBR_L1)) on L1 Cache. Furthermore we have analyzed various pairs of algorithms on L1 and L2 respectively, resulting in finding a suitable pair of replacement algorithms. Simulation on L1 shows, among the considered existing replacement algorithms FIFO is performing better than others. While the proposed replacement algorithm PBR_L1 is working about 1.7% to 44 % better than FIFO for various cache sizes. The analysis for various pairs on L1 and L2 respectively shows that among the considered existing various pairs the best pair is FIFO followed by FIFO. While the proposed replacement policy PBR_L1 followed by FIFO works approximately 66% to 100% better than the pair FIFO-FIFO for various cache sizes. Furthermore simulation results by fixing the cache size L1 and L2 and varying list length shows that the performance of proposed algorithm on L1 is better than others considered in this paper. Similar analysis done for various pairs shows that the pair PBR_L1 on L1 followed by FIFO on L2 is superior to other pairs for varying length list.

—————————— ◆ ——————————

## 1 INTRODUCTION

Cache memory is used for speed matching of main memory and processor. Cache memory works with the principle of locality. The principle of locality refers that CPU does not requires all the code/data at a time. The principle of locality can be spatial or temporal [1, 2, 3].

Whenever a page/word/block is requested from CPU, first of all it is searched on L1 if the required page is found in L1 it is a hit else a miss. When L1 is saturated and it is a miss then a block from L1 is to be evicted to create a space for the required page. Various replacement algorithms, such as LRU, FIFO, LFU [4, 5] etc are used to select the victim page.
L1 is having better temporal locality than L2, as L2 is accessed when a miss occurs on L1.
Hot pages should remain in L1 and cold pages should be taken off and are placed in the main memory.
Whenever a page is evicted from L1 it will be placed on L2. The nature of pages which should reside on L2 should be neither too cold nor too hot i.e. moderate. Place for hot pages is L1 and that for cold pages is in the main memory [3]. Most of the algorithm tries to keep hot pages in the cache but from the above discus-sion it is clear that this is not the requirement for L2 cache. Thus the replacement algorithm which is suitable on L1 may not be suitable on L2.
Furthermore various algorithms and applications such as Matrix Multiplication, Fast Fourier transform, Networks, N Databases etc., will have varying accesses to the memory thus resulting in varying principal of locality. Initially if any word/block/page is referenced then it will suffer with compulsory miss. If a reference suffers a miss because of saturated cache or capacity miss, then replacement algorithm will evict a word/block/page. The replacement algorithms are based on some criteria may be recency, frequency etc. The replacement algorithms which are taken into account are Least Recently Used (LRU), First in First out (FIFO), Least Frequently Used (LFU).

Based on above discussion this paper has proposed a new replacement policy PBR-L1 on L1, which takes the benefit of access pattern of Merge Sort and thus works better than other existing replacement policies considered. Besides this, we have also exerted to find a proficient pair of replacement algorithms on L1 and L2 respectively.

## 2 RELATED WORK

Most of the studies for replacement policies have been done for L1. Replacement algorithms can be recency based or frequency based or may follow both the aspects,

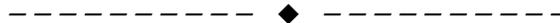

- *Richa Gupta is with the Department of Computer science and Information Technology, KCB Technical Academy, Indore.*
- *Sanjiv Tokekar is with the Department of Electronics and telecom engineering, IET, DAVV, Indore.*

such as Least Recently Used (LRU) [4, 5].Least Recently Used–K (LRU-K) [6], Most Recently Used (MRU) [4,7] Least Frequently Used (LFU) [4,7] Least Frequently Used-K (LFU-K) [7,8].Least Frequently Recently Used (LFRU) [9] Low Inter-reference Frequency (LIRs) [10], 2Q [11], Second Chance Frequency- Least Recently Used (SF-LRU) [12] etc. Ismail Ari *et.al* suggests an adaptive Caching Scheme using Multiple Experts (ACME). It proposes the use of machine learning algorithms to select the current best policy or mixture of policies by allowing each adaptive cache node to tune itself based on work load it observes [13]. Much work has not been done for L2 cache replacement algorithms. A comprehensive study of second level cache management was given by Zhou *et.al* [14]. which emphasizes that access pattern of second level cache is different than first level. More specifically it presents a new algorithm Multi Queue (MQ) to effetievly manage second-level buffer caches, which were basically designed for single level. Michael *et.al* proposes a policy Karma which uses application hints to partition the cache and to manage each range of blocks with the policy best suited fot its access pattern [15] ayne *et.al* presents a replacement policy based on the detection of temporal locality in the l2 cache, the block to be taken out is chosen by considering both its priority in the LRU stack and whether it exhibits temporal locality or not [16]. uhui Li *et.al* illustrates how the information contained in writes from the first tier can be used to improve the performance of the second tier [17].

It has been discussed earlier that the same replacement algorithm may not be suitable for various applications and algorithms, which are having varying principle of locality. This paper discusses about the access pattern of Merge Sort. Zhang *et.al* has focused on reducing I/O time during merge phase for external merge sort [18]. LaMacra *et.al* had explored that cache conscious design and analysis in classical sorting algorithms leads to potential performance gain. [19].Brodal *et.al* had done a detailed experimental study of cache oblivious sorting algorithms [20]. Franceschini showed how to perform optimal cache oblivious sorting implicitly using only O(1) space[21]. Juszczak had shown an efficient implementation of merge sort. It is based on a fast half copying merge algorithm [22]. Zhang *et.al* had applied dynamic memory utilization in sorting [23]. Xiao *et.al* has discussed about considering memory hierarchy for sorting algorithms at the time of design and implementation [24].

## 3 REPLACEMENT ALGORITHMS ON L1

Initially if any word/block/page is referenced then it will suffer with compulsory miss. If a reference suffers a miss because of saturated cache i.e. capacity miss, then replacement algorithm will evict a word/block/page. The replacement algorithms which are taken into account are Least Recently Used (LRU), First in First out (FIFO), Least Frequently Used (LFU).

## 4 PROPOSED REPLACEMENT ALGORITHM ON L1

Merge sort works on the principle of divide and conquer. The list is divided into two halves, and then first half is again divided into two halves and so on till there are two elements in the list. These two elements are sorted. Now merging following the sorting order is done between the small lists. This merging sorting and merging process goes on till we obtain a sorted list for first half. The same process is done for second half, which results in sorted list for second half. Now again merging following the sorting order is done between these two sorted lists. After analyzing the access pattern of merge sort for various length of lists. We have noticed that in merge sort the reference pattern can be divided in three categories. First and second category gives the references of sorted list for first half and second half respectively, the third category of the reference merges these two sorted list.

While developing new algorithm on L1 we have emphasized our efforts on these three categories. The idea is to partition the cache in two parts P1 and P2 as shown in Fig. 1, very small part for fixed cache (P1) and a large part for variable cache (P2). In the fixed part of the cache the replacement takes place for only for selected few elements. For the first category of references, after initial misses first few elements use the fixed part cache. Similarly for the second category of reference, the same fixed part cache is used for last few references as now there will be no reference from first part. Now for the reference of third category the rest of the cache i.e. the variable part cache is used for merging the two sorted lists. Thus for L1 we have developed Partition Based Replacement (PBR-L1) Algorithm.

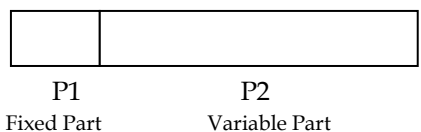

Fig. 1. Cache divided in two parts Fixed Part and Variable Part

The proposed replacement policy PBR_L1 is compared with the policies LRU, FIFO and LFU on L1 for various cache sizes.

## 5 REPLACEMENT ALGORITHMS ON L1 AND L2

As discussed earlier it has been realized that L2 is having poor temporal locality as compared to L1. If the required page is not in L1 it is searched in L2 and then in main memory. It means that L2 will also suffer with initial misses as L1. As reference to L2 is less frequent than to L1 and L2 is greater than L1, so after the initial misses the probability of the data to remain in L2 is high. As the locality of reference is different for L1 and L2 the algorithm which is suitable for L1 may not fit for L2.

After various simulation results it has been noticed that the replacement algorithms when paired with LRU or LFU; its performance is same. Thus we are showing the results of any one pair out of these.

## 6 PERFORMANCE ANALYSIS

### 6.1 Performance Analysis of PBR-L1 on L1

To analyze the behavior of the replacement algorithms mentioned in section 3 reference pattern of merge sort is generated. For simulation length of the list considered is 256. For each replacement algorithm, Miss Rate is calculated for varying size of L1. With the variation of size of the list the required maximum size of L1 will vary.

The results of various replacement algorithms on L1 are as shown in Fig. 2, which gives the miss rates for different cache sizes.

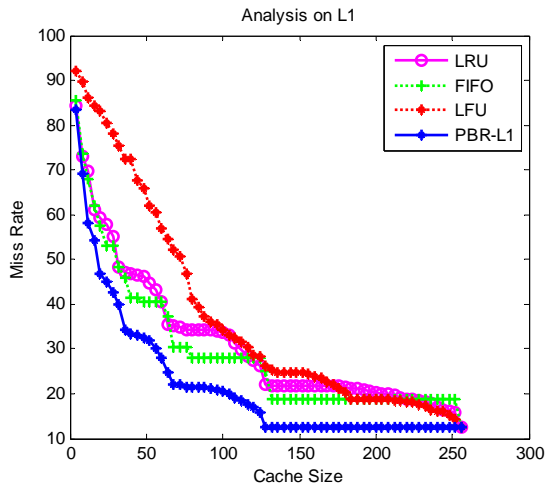

Fig. 2. Comparison of PBR-L1 with replacement algorithms LRU, FIFO, LFU on L1.

From Fig. 2 it can be depicted that among the considered existing algorithm FIFO is performing better than LRU and LFU. It can be clearly seen with the help of Fig. 2 that the proposed replacement policy PBR_L1 is the best. The performance of the proposed replacement algorithm PBR_L1 is ranging from 1.7% to 44% better than FIFO for various cache sizes.

### 6.2 Performance Analysis of Pair of Algorithms on L1 and L2

For analyzing the behavior of various combinations of replacement algorithms on L1 and L2, size of L1 is fixed and size of L2 is varied. The following pairs are being analyzed

CASE I: Replacement algorithm on L1 is LRU, while on L2 the replacement algorithms LRU, FIFO, LFU are applied.

CASE II: Replacement algorithm on L1 is FIFO and LRU, FIFO, and LFU on L2 for the same length of list as discussed above.

CASE III: The proposed replacement algorithm PBR-LRU is used on L1 and LRU, FIFO, and LFU are used on L2.

The results of CASE I, II and III are as shown in Fig.3, Fig. 4, Fig. 5 respectievely.

In CASE I the comparison is donr for the pairs LRU-LRU; LRU-LFU; LRU-FIFO. With the help of Fig. 3, it can be realized that the pair LRU-FIFO is giving better results than the other two pairs.

In CASE II the comparison is done for the pairs FIFO-LRU; FIFO-LFU; FIFO-FIFO. With the help of Fig. 4, it can be realized that the combination FIFO- FIFO is working better than other pairs.

In CASE III the proposed algorithm PBR_L1 is applied on L1. Here we are comparing the algorithm PBR_L1 on L1 followed by LRU, LFU and FIFO on L2. The results for these pairs are as shown in Fig. 5. With the help of Fig. 5, it can be realized that the pair PBR_L1-FIFO is performing much better than other two pairs.

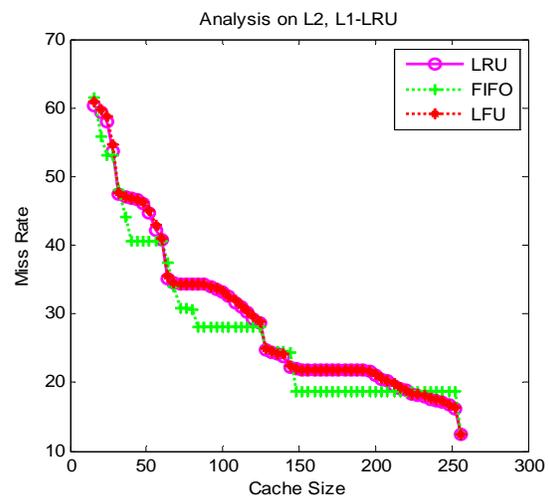

Fig. 3. (L1 Size: 8, L2 Varied from 16) L1-LRU

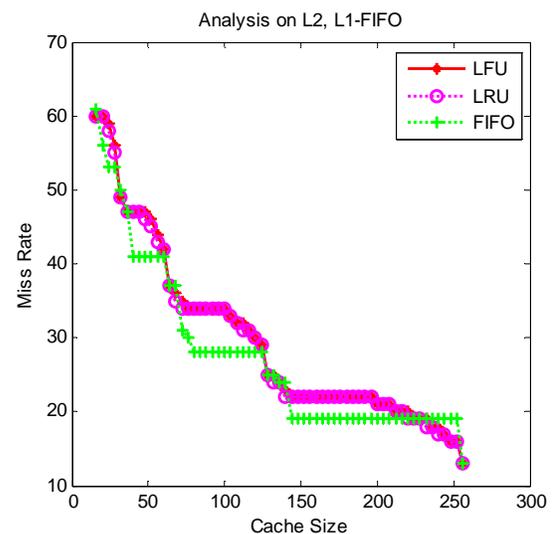

Fig. 4. (L1 Size: 8, L2 Varied from 16) L1-FIFO

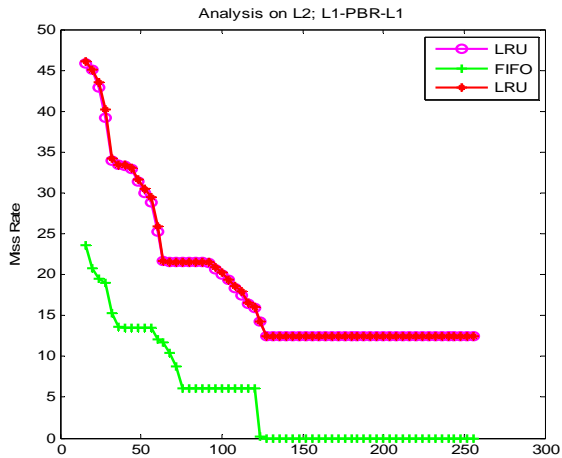

Fig. 5. (L1 Size: 8, L2 Varied from 16) L1-PBR-L1

The results of the analysis illustrated above are combined in tabulation form as shown in Table 1. With the help of Table it can be explored that the pair PBR_L1-FIFO is superior to other pairs. This pair is working approximately 66% to 100% better than all other pairs for various cache sizes.

TABLE 1
COMPARISION OF VARIOUS PAIRS OF REPLACEMENT ALGORITHMS

| ALGORITHM L1 | L2 | CACHE SIZE 16 | 56 | 96 | 136 | 176 | 216 | 256 |
|---|---|---|---|---|---|---|---|---|
| LRU | LRU | 60.89 | 43.21 | 33.89 | 21.88 | 21.73 | 19.14 | 12.50 |
|  | FIFO | 61.96 | 40.63 | 28.13 | 18.75 | 18.75 | 18.75 | 12.50 |
|  | LFU | 61.23 | 43.85 | 34.03 | 21.88 | 21.83 | 19.24 | 12.50 |
| FIFO | LRU | 60.30 | 43.12 | 33.74 | 21.88 | 21.88 | 19.43 | 12.50 |
|  | FIFO | 61.52 | 40.63 | 28.13 | 23.97 | 18.75 | 18.75 | 12.50 |
|  | LFU | 60.79 | 43.75 | 33.84 | 21.88 | 21.88 | 19.63 | 12.50 |
| PBR_L1 | LRU | 46.68 | 28.86 | 20.12 | 12.50 | 12.50 | 12.50 | 12.50 |
|  | FIFO | 24.07 | 13.48 | 6.05 | 0.00 | 0.00 | 0.00 | 0.00 |
|  | LFU | 47.02 | 29.25 | 20.17 | 12.50 | 12.50 | 12.50 | 12.50 |

## 7 FIXING CACHE SIZE AND VARYING LENGTH OF LIST

An additional criterion for the analysis is taken as fixing the size of L1 and L2 and varying the list size and evaluating the miss rate. Here we are fixing the size of L1 to 32, L2 to 128 and varying the list size from 8 to 1024. This analysis compares the performance of PBR-L1 on L1 with other replacement algorithms. The comparison of the proposed algorithm PBR_L1 with others is as shown in Fig. 6. for varying List Length. Fig. 7 compares the various pair of replacement algorithms PBR_L1 on L1 followed by LRU, LFU and FIFO on L2.
From the Fig. 6 it can be realized that for almost for all the list size the proposed replacement policy PBR_L1 is performing much better than others. With the help of Fig. 7 it can be analyzed that overall the performance of the pair PBR_L1 on L1 and FIFO on L2 is better than other pairs for maximum list sizes.

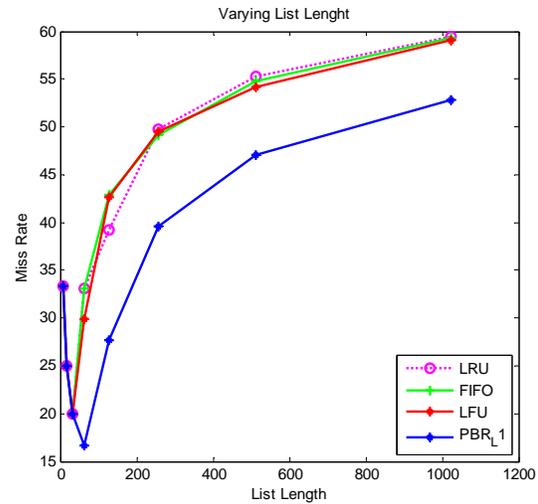

Fig. 6. Analysis for varying List Length on L1

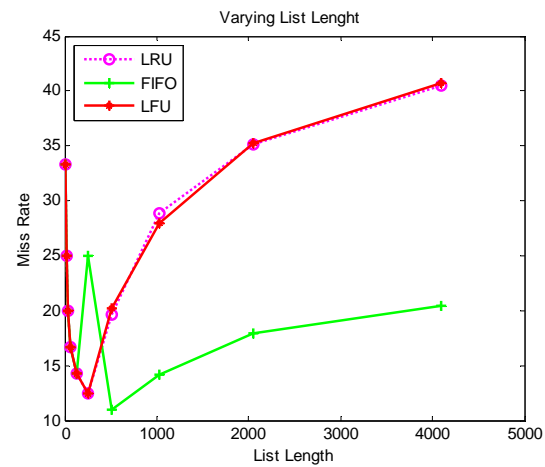

Fig. 7. Analysis for Varying List Size for the pair; on L1 is PBR_L1 followed by LRU, LFU, FIFO.

## 8 CONCLUSION

This paper is basically concentrated to explore the efficient pair of page replacement algorithms on L1 and L2 respectively, if the memory reference string is of merge sort. With help of simulation result it has been realized that the replacement policy FIFO is performing better than the replacement policies LRU and LFU. After analyzing the access pattern of merge sort, we have proposed a new replacement algorithm PBR_L1 which works about 1.7% to 44% better than the FIFO replacement algorithm for various cache sizes. Along with this we have also exerted to disclose a pair of replacement policies which are paramount on L1 and L2 respectively for the same application. When we compare the various pairs for replacement policies, the simulation shows that the pair of proposed algorithm PBR_L1 on L1 and FIFO on L2 is superior to other pairs considered in this paper. This pair is performing about 66% to 100% better than the pair FIFO-FIFO for various cache sizes. Furthermore when we fix the cache size and vary the list length then on L1 the performance of PBR_L1 is better than others. While the same analysis done for the pair of algorithms shows that the pair PBR_L1 on L1 and FIFO on L2 is performing better than other pairs.

## Authors Profile

**Richa Gupta** has received her M.Tech degree in computer science in the year 2004 fron Devi Ahilya Vishwavydyalaya, Indore. Currently she is pursuing her Ph.D. in computer engineering from Institute of Engineering and Technology, Devi Ahilya University, Indore. She is having around 13 years of teaching experience and around 3 years of research experience. Currently she is working as Reader in Computer science and Information Technology Department at KCB Technical Academy. Her subject of interest includes Computer Architecture, Operating System, Parallel Processing and Digital Computer Organization. Currently she is pursuing her Ph. D. under guidance of Dr. Sanjiv Tokekar, Professor, IET, DAVV, Indore. She has published research papers in National/ International Conferences and Journals. Her research area is in the field of replacement algorithms for L1 and L2 cache.

**Sanjiv Tokekar** has completed his BE in Electronics in the year 1982 and has completed his M.E in Electronics specialization in Applied Electronics in the year 1985. He has completed his Doctoral Degree in the year 1996 in Electronics Engineering.He has 27 years of teaching experience. Presently he is working as professor and Head in Electronics and Telecommunication engineering Departmnet, IET, DAVV. Indore, India. Under his guidance 5 researchers have successfully compelted their research. He has published 60 papers 7 in International Journal, 2 in national journal, 44 in international conferences and 6 in inational conferences. He is a senior member of IEEE, life member of CSI, life member of ISTE. He has been invited as session chair in National and Internatioanal Conferences.